\documentclass[]{interact}

\usepackage[caption=false]{subfig}% Support for small, `sub' figures and tables

\usepackage[lined,boxed]{algorithm2e}

\usepackage[numbers,sort&compress]{natbib}% Citation support using natbib.sty
\bibpunct[, ]{[}{]}{,}{n}{,}{,}% Citation support using natbib.sty
\makeatletter% @ becomes a letter
\def\NAT@def@citea{\def\@citea{\NAT@separator}}% Suppress spaces between citations using natbib.sty
\makeatother% @ becomes a symbol again

\theoremstyle{plain}% Theorem-like structures provided by amsthm.sty

\theoremstyle{definition}

\theoremstyle{remark}

\begin{document}

%\articletype{ARTICLE TEMPLATE}% Specify the article type or omit as appropriate

\title{Estimating a pressure dependent thermal conductivity coefficient with applications in food technology}

\author{
\name{Marcos A. Capistr\'an\textsuperscript{a}\thanks{CONTACT Marcos A. Capistr\'an. Email: \textit{marcos@cimat.mx}} and Juan Antonio Infante del R\'io\textsuperscript{b}}
\affil{\textsuperscript{a}Centro de Investigaci\'on en Matem\'aticas (CIMAT), Jalisco S/N, Valenciana, Guanajuato,
36023, M\'exico; \textsuperscript{b}Instituto de Matem\'atica Interdisciplinar and Departamento de Matem\'atica Aplicada,
Facultad de CC. Matem\'aticas, Universidad Complutense de Madrid, Plaza de Ciencias 3,
28040, Madrid, Spain}
}

\maketitle

\begin{abstract}
In this paper we introduce a method to estimate a pressure dependent thermal conductivity
coefficient arising in a heat diffusion model with applications in food technology.
To address the known smoothing effect of the direct problem, we model the uncertainty of the conductivity 
coefficient as a hierarchical Gaussian Markov random field (GMRF) restricted to uniqueness conditions for the 
solution of the inverse problem established in Fraguela {\it et al.}~\cite{fraguela2013uniqueness}. 
Furthermore, we propose a Single Variable Exchange Metropolis-Hastings algorithm to sample the 
corresponding conditional probability distribution of the conductivity coefficient given observations of the 
temperature. Sensitivity analysis of the direct problem suggests that large integration times are necessary
to identify the thermal conductivity coefficient. Numerical evidence indicates that
a signal to noise ratio of roughly $10^{3}$ suffices to reliably retrieve the thermal conductivity
coefficient.
\end{abstract}

\begin{keywords}
Metropolis-Hastings, Food technology, Thermal conductivity coefficent, Gaussian Markov Random Field, 
Hierarchical model
\end{keywords}

\section{Introduction}
\label{sec:intro}

In this paper we introduce a method to estimate a time dependent thermal conductivity
coefficient arising in a heat diffusion model with applications in food technology. Therefore,
we are interested in an inverse problem governed by a linear parabolic partial
differential equation. According to Infante et al.
~\cite{infante2009modelling} ``...{\it In high-pressure processes, food is treated with mild temperatures
and high pressures in order to inactivate enzymes and preserve as many of its organoleptic properties
as possible}...". Consequently, there are several research efforts to model the dynamics of food
preservation processes which rely on high pressures away from thermal equilibrium
\cite{adams1991enzyme,cardoso1978new,hendrickx1998effects,denys2000modeling,
otero2007model,norton2008recent,infante2009modelling,
otero2010modeling,Smith2014,serment2014high}. A related problem is to infer the thermal conductivity
coefficient associated to food preservation processes given measurements of the temperature, see 
\cite{infante2015identification}. Of particular importance towards the solution of the inverse problem 
are results on the structural identifiability of the thermal conductivity coefficient, see \cite{fraguela2013uniqueness}.

On the other hand, inferring a thermal conductivity coefficient from temperature observations is a statistical 
inference problem. Indeed, it is necessary to model first a forward mapping defined by a well posed problem
for a partial differential equation with initial and boundary conditions (e.g. Fraguela et al.
\cite{fraguela2013uniqueness} give conditions for a mapping from thermal conductivity coefficient
to temperature to be well defined). Secondly, we must postulate an observation mapping from state variables
to data taking into account a signal to noise ratio. This double tier modeling renders the inverse
problem as a statistical inference problem. Related previous work includes boundary heat flux and heat 
source reconstruction in heat conduction problems using the Bayesian paradigm by Wang and Zabaras~\cite{wang2004bayesian,wang2004hierarchical}. More in general, Kaipio and 
Fox~\cite{kaipio2011bayesian} offer a review of results and challenges in the Bayesian solution
of heat transfer inverse problems.
We remark that inverse heat conduction problems are difficult given 
the smoothing nature of the diffusion operator, see Isakov \cite{isakov2006inverse},
giving rise to a situation where careful modeling and a large signal to noise ratio is necessary in order 
to be able to solve the inverse problem in a practical setting. Of note, thermocouples, which 
are data acquisition devices in the setting considered here, have a signal to noise ratio of roughly $10^{3}$
within all the working temperature regime. 

In this paper, we rely on the Bayesian paradigm to model the prior
information about the parameter to be inferred. The rationale is as follows. In the solution of
inverse problems defined by partial differential equations, it is of paramount importance to
model prior distributions that are informed with theoretical features of the governing differential
equation. On the other hand, Zellner
\cite{zellner1988optimal} shows that if Bayes theorem is regarded as a learning process in the context of
information theory, then the amount of input information is preserved into the output information coded in the
posterior distribution. Consequently, in this paper, we construct a prior distribution that incorporates Theorem 14
of~\cite{fraguela2013uniqueness}, giving rise to an unimodal posterior distribution. We analyze the arising
posterior distribution using a specially tailored variant of the Metropolis-Hastings algorithm.

The paper is organized as follows: In Section \ref{sec:intro} we describe the direct problem that defines
the forward mapping, as well as all theoretical results necessary to pose the two tier formulation of the
inverse problem as a statistical inference problem. In Section \ref{sec:res&disc} we show our results and
discuss our findings. Finally, in Section \ref{sec:concl} we reflect upon the reaches and limitations
of our approach and offer some perspectives.

\section{Problem statement}
\label{sec:intro}

We shall consider the mathematical modeling of a food preserving method based on high pressure processing
away from thermal equilibrium, see Infante {\it et al.}~\cite{infante2009modelling}. The physical system is
partially observed, thus it gives rise to an inverse problem where we want to infer a thermal conductivity
coefficient given temperature observations. More precisely, the thermal conductivity coefficient depends on
pressure only, e.g. $k=k(P)$, see \cite{fraguela2013uniqueness}. The overall goal is identifying $k(P)$ from
measurements of the temperature $T$. Of note, data aquisition design renders pressure $P$ as a known
and strictly increasing function of time. Therefore, the problem of identifying $k(P)$  is equivalent to identify
$k(t)=k(P(t))$.

\subsection{Direct problem or forward mapping}
\label{sec:dir_prob}
According to Smith {\it et al.} \cite{Smith2014}, a mathematical model of the high pressure food preserving
method under consideration can be cast as an initial and boundary value problem for a parabolic partial
differential equation as follows
\begin{equation}
\label{eq:pb_unid_1}
\left\{\begin{array}{ll}
  \varrho C_{p} \dfrac{\partial T}{\partial t}-k(t)\Delta T=\alpha \dfrac{\rm{d} P}{\rm{d} t}(t) T & \mbox{ in }\:B_{R}\times (0,t_{\rm f})\\[3mm]
  k(t)\dfrac{\partial T}{\partial {\bf n}}=h\left(T^{e}(t)-T\right)
  & \mbox{ on $\partial B_{R}\times (0,t_{\rm f})$}\\[3mm]
  T=T_{0} & \mbox{ on }\:B_{R}\times \{0\},
\end{array}
\right .
\end{equation}
where $B_R\subset R^2$ is a disk with center $(0,0)$ and radius $R>0$, and $t_{\rm f}>0$ is
the final time. Other model components are as follows: $\alpha>0$ is the coefficient of thermal expansion,
$\varrho >0$ the density and $C_p>0 $ the specific heat; $P\in {\cal C}^{1}([0,t_{\rm f}])$ is the
pressure at time $t$; $k(t)\geq k_0>0$ is the thermal conductivity; $T^{e}$ is the external temperature;
$\mathbf{n}$ is the outward unit normal vector at the boundary of $B_R$; $h>0$ is a heat exchange
coefficient and $T_0$ is the initial temperature (assumed to be constant).

Other conditions being equal, problem~(\ref{eq:pb_unid_1}) has a unique (classical) radial
solution $T$, and defines a {\it forward mapping}
\begin{equation}
\label{eq:forward_mapping}
F(k)=T,
\end{equation}
i.e., given a thermal conductivity coefficient $k=k(t)$, it is possible to evaluate a unique temperature $T$.
In the context in inverse problems, we care about conditions on equation~\eqref{eq:forward_mapping}
to render parameter $k$ {\it uniquely identifiable} given data $T$. We shall summarize known uniqueness results
for the inverse problem in the next section and use them to pose the inverse problem as a statistical
inference problem.

\subsection{Inverse problem}
\label{sec:inv_prob}

The inverse problem that we will consider in the paper is to estimate the thermal conductivity coefficient
$k$ given experimental measurements of the temperature $T$. In order to ensure the uniqueness of
function $k$, the following context is established in Fraguela {\it et al.}~\cite{fraguela2013uniqueness}:
Temperature is measured at two points, one of them on the boundary of the domain
and the other
one at distance $r_0$ of the center point; also, the following hypotheses are assumed to hold
\begin{enumerate}%[{\bf (H1)}]
\item [{\bf H1}] $T^{e}(t)\equiv T_{0}$ for all $t\in[0,t_{\rm f}]$.
\item [{\bf H2}] $P$ is a linear function in $[0,t_{\rm f}]$ with $\frac{\rm{d} P}{\rm{d} t}\equiv\beta>0$  and $P(0)$ reaches the atmospheric
pressure value (therefore, $k(0)$ is known).
\item [{\bf H3}] $k$ is a right locally analytic function in $[0,t_{\rm f})$.
\item [{\bf H4}] \begin{displaymath}
\int_{0}^{t_{\rm f}} k( t )dt \le \varrho C_{p} \frac{(R-r_{0})^{2}}{4}.
\end{displaymath}
\item [{\bf H5}] \begin{equation}\label{eq:supp0}
\frac{\rm{d} k}{\rm{d} t} (t)\le\frac{\alpha\beta}{\varrho C_{p}}\frac{k(t)}{e^{\frac{\alpha\beta}{\varrho C_{p}} t}-1},\:t\in[0,t_{\rm f}].
\end{equation}
\end{enumerate}
Note that hypothesis {\bf (H5)} means an upper estimate of the logarithmic derivative of $k$. Then, taking in account the positivity of $k$, we write
\begin{equation}
\label{eq:kexpu}
k(t)=\exp(u(t)).
\end{equation}
In order to discretize $u$, we shall use an equidistant grid on interval $[0,t_{\rm{f}}]$
\begin{displaymath}
t_{j}= j\tau,~j=0,1,...,n,~n\in\mathbb{N},~\tau=\frac{t_{\rm{f}}}{n}
\end{displaymath}
and the finite element basis of piecewise linear functions $\{\varphi_{i}\}_{i=0}^{n}$ on $[t_{j},t_{j+1}], i=0,1,\ldots,n-1$ such that $\varphi_{i}(t_{j})=\delta_{ij}$ (Kronecker delta). We assume that
\begin{equation}
\label{eq:discrete_k}
u(t)=\sum_{i=0}^{n}u_{i}\varphi_{i}(t)
\end{equation}
and we want to determine $u_{i} =u(t_i)$ for $i\geq1$ ($u_{0}=\log(k(0))$ is known).

From \eqref{eq:discrete_k}, hypothesis {\bf (H4)} can be discretized as follows
\begin{equation}\label{eq:supp1}
\tau\sum_{i=0}^{n-1}\frac{\exp(u_{i+1})-\exp(u_{i})}{u_{i+1}-u_{i}} \leq \varrho C_{p}
\frac{(R-r_{0})^{2}}{4}.
\end{equation}

On the other hand, in order to increase the robustness of our approximations, we weaken restriction {\bf (H5)}
posing it in variational form. First, if we write \eqref{eq:supp0} in terms of $u$ we have
\begin{equation}\label{eq:supp0u}
\frac{\rm{d} u}{\rm{d} t}(t)\le f(t),\:t\in[0,t_{\rm f}].
\end{equation}
where $\displaystyle{f(t)=\frac{\alpha\beta}{\varrho C_{p}}\frac{1}{e^{\frac{\alpha\beta}{\varrho C_{p}} t}-1}}$.

Now, multiplying equation \eqref{eq:supp0u} by $\varphi_{i}(t), i=1,2,\ldots,n$, integrating by parts and using Simpson's rule to approximate the definite integrals, hypothesis {\bf (H5)} becomes
\begin{equation}\label{eq:supp2}
\frac{u_{i+1}-u_{i-1}}{2} \leq \frac{\tau}{3} \left( f\left(\frac{t_{i}+t_{i-1}}{2}\right)+f(t_{i})+f\left(\frac{t_{i+1}+t_{i}}{2}\right) \right),
\end{equation}
for $i=1,2,\ldots,n-1$. Theorem 14 in \cite{fraguela2013uniqueness} establishes that the inverse problem has
an unique solution if the temperature $T$ is solution of problem \eqref{eq:pb_unid_1} and hypotheses
{\bf (H1)}-{\bf (H5)} hold.
Let us introduce the following notation. Let us denote by $Q$ the set of functions $u$
such that the corresponding $k(t)$ in equation~\eqref{eq:kexpu} satisfies $k_{i}\ge k_{0}, i=1,2,\ldots,n$
and hypotheses {\bf (H1)}-{\bf (H5)} hold, i.e., $u$ satisfies equations~\eqref{eq:supp1} and~\eqref{eq:supp2}.
In Subsection \ref{sec:stat_inv} we shall use the above notation to introduce a two tiered formulation
of the inverse problem.

\subsection{Statistical inversion}
\label{sec:stat_inv}

In this Subsection we shall use the optimality property of Bayes formula as a learning process, see Zellner \cite{zellner1988optimal}, to incorporate Fraguela et al \cite{fraguela2013uniqueness} uniqueness
theorem into our prior model of the thermal conductivity coefficient, which we discretize as
equation~\eqref{eq:discrete_k}. Next, we shall correct the prior model with a likelihood function
that depends on both, the forward mapping~\eqref{eq:forward_mapping} and data. Finally, we propose
a variant of the Metropolis-Hastings algorithm to explore the arising posterior distribution.

For the sake of clarity, throughout the current Subsection we shall use bold uppercase letters to denote random
variables, while instances of the random variables are denoted as in the direct problem.

Let ${\bf \hat{U}}=({\bf U}_{0},...,{\bf U}_{n})$ denote the coefficients in equation~\eqref{eq:discrete_k} and
${\bf\Sigma}_{2}$ its standard deviation, which we assume is the same for all ${\bf U}_{i}$, $i=0,...,n.$

We model the prior probability density of ${\bf \Theta}=({\bf \hat{U}},{\bf \Sigma}_{2})$ using a hierarchical
strategy as follows
\begin{equation}
\label{eq:prior1}
\pi_{\bf\Theta}(\theta)=
\pi_{\bf\Theta}(\hat{u},\sigma_{2})
=\pi_{{\bf \hat{U}}|{\bf \Sigma}_{2}}(\hat{u}|\sigma_{2})\pi_{{\bf \Sigma}_{2}}(\sigma_{2})\chi_{Q}(\hat{u}),
\end{equation}

where we denote $\hat{u}=(u_{0},u_{1},\cdots,u_{n})^{T}$. First, we define the indicator function
\begin{displaymath}
\chi_{Q}(u)=
\begin{cases}
1&\text{if $u(t)=\sum_{i=0}^{n}u_{i}\varphi_{i}(t)$ satisfies~\eqref{eq:supp1},~\eqref{eq:supp2}
and $u_{i}\ge u_{0}=\log(k(0))$},\\
0 &\text{otherwise.}
\end{cases}
\end{displaymath}

Next, we model the prior distribution of $\sigma_{2}$ using a Gamma distribution with probability
density function
\begin{displaymath}
\pi_{\bf{\Sigma_{2}}}(\sigma_{2})=\frac{1}{\Gamma(a)b^{a}}\sigma_{2}^{a-1}
\exp\left(-\frac{\sigma_{2}}{b}\right).
\end{displaymath}
The rationale is that both, the Gamma distribution and the hyperparameter $\sigma_{2}$ have support
on the positive real numbers, and the Gamma distribution is characterized by two instrumental parameters
$a$, $b$, i.e., $\mathbb{E}(\sigma_2)=ab$, ${\rm var}(\sigma_{2})=ab^{2}$. In the examples below we choose
$a=b=1$ such that the prior distribution is monotonically decreasing.

Now, in order to model $\pi_{\bf \hat{U}|\Sigma_{2}}(\hat{u}|\sigma_{2})$, let us consider the following tridiagonal
matrix $\hat{A}\in\mathbb{R}^{(n+1)\times (n+1)}$
\begin{displaymath}
\hat{A}=\frac{1}{\tau^{2}}\left(\begin{array}{ccccc}
         2 & -1 & & &  \\
         -1 & 2 & -1 &  & \\
         & \ddots & \ddots & \ddots & \\
         & & -1 & 2 & -1 \\
         & & & -1 & 2
        \end{array}
  \right),
\end{displaymath}

where $\tau=\frac{t_{\rm{f}}}{n}$ as before. Of note, $\hat{A}$ is a symmetric positive definite matrix
arising if we discretize the negative Laplacian operator with Dirichlet boundary conditions using centered
finite differences.
Let us denote $\hat{\Sigma}=\hat{A}^{-1}$. Then, according to Bardsley~\cite{bardsley2013gaussian},
the expression
\begin{displaymath}
\det(2\pi\hat{\Sigma})^{-1/2}\exp\left(-\frac{1}{2}\hat{u}^{T}\hat{A}\hat{u}\right)
\end{displaymath}
defines a Gaussian Markov Random Field (GMRF) with precision matrix $\hat{A}$
(and, covariance matrix $\hat{\Sigma}$). Moreover, we must condition this GMRF to the fact that the first value $u_{0}$ is known. Let us consider $(1,n)$--block
partition of matrices $\hat{\Sigma}$ and $\hat{A}$
\begin{displaymath}
\hat{\Sigma}=\left(\begin{array}{cc}
         \hat{\Sigma}_{11} & \hat{\Sigma}_{12} \\
         \hat{\Sigma}_{21} & \hat{\Sigma}_{22} \\
        \end{array}
  \right)
,\;
\hat{A}=\left(\begin{array}{cc}
         \alpha & v^{T} \\
         v & A \\
        \end{array}
  \right).
\end{displaymath}

Then, our conditional density can be taken as a multivariate normal $N(\mu,\Sigma)$, where
the mean $\mu$ is the vector $u_{0}\hat{\Sigma}_{21}\hat{\Sigma}_{11}^{-1}$ and the covariance
matrix $\Sigma$ is the Schur complement of matrix $\hat{\Sigma}_{22}$ in $\hat{\Sigma}$, i.e.
\begin{displaymath}
\Sigma=\hat{\Sigma}_{22}-\hat{\Sigma}_{21}\hat{\Sigma}_{11}^{-1}\hat{\Sigma}_{12}.
\end{displaymath}

Now, letting $u=(u_{1},...,u_{n})$ we are ready to set
$\pi_{\bf \hat{U}|\Sigma_{2}}(\hat{u}|\sigma_{2})=\frac{h(u,\sigma_{2})}{Z(\sigma_{2})}$
where
\begin{displaymath}
h(u,\sigma_{2})=\det(2\pi\Sigma)^{-1/2}\exp\left(-\frac{1}{2}(u-\mu)^{T}A(u-\mu)\right),
\end{displaymath}
and  $Z_{0}(\sigma_{2})$ is some unknown normalization constant.

Therefore, our prior probability density in equation \eqref{eq:prior1} can be written as
\begin{equation}
\label{eq:prior2}
\pi_{\bf\Theta}(\theta)=\frac{h(u,\sigma_{2})}{Z_{0}(\sigma_{2})}\pi_{\bf\Sigma_{2}}(\sigma_{2})
\chi_{Q}(u).
\end{equation}

{\bf Remark.} The single variable exchange variant of the Metropolis-Hastings algorithm described below
does not require the knowledge of $Z_{0}(\sigma_{2})$ although it depends on $\sigma_{2}$.

Next, we derive a formula to evaluate the likelihood of data $T$ given an instance of the parameter
$u=(u_1,...,u_n)$, assuming $u_{0}=\log(k(0))$ is known. Let us denote
$G(u)_{j}=\exp\left(\sum_{l=0}^{n}u_{l}\varphi_{l}(t_j)\right)$. We assume that data satisfies

\begin{equation}
\label{eq:regression}
T_{ij}=F(G(u)_{j})_{i}+\xi_{ij}
\end{equation}

\noindent where data $T_{ij}=T(r_{i},t_{j})$ is a measurement of the temperature at points $r_{i}$
and times $t_{j}$, while $F\left(G(u)_{j}\right)_{i}$ is the forward
mapping \eqref{eq:forward_mapping} acting on a instance
$u$ of the parameters, while $\xi_{ij}$ is Gaussian noise
with mean $0$, standard deviation $\sigma_{1}$ and probability distribution
$\xi_{ij}\sim\mathcal{N}(0,\sigma_{1}^{2})$. Under the hypothesis of independence of the $\xi_{ij}$, we
can approximate the conditional probability distribution $\pi_{{\bf T|\Theta}}(T|\theta)$ of $T$ by
the product

\begin{equation}
\label{eq:likelihood}
\pi_{{\bf T|\Theta}}(T|\theta)=
\prod_{i,j}\frac{1}{(\sqrt{2\pi}\sigma_{1})^{2n}}\exp\left(-\frac{1}{2\sigma_{1}^{2}}
\left(T_{ij}-F\left(G(u)_{j}\right)_{i}\right)^{2}\right)
\end{equation}

Likelihood~(\ref{eq:likelihood}) is equivalent to
$T_{ij}\sim\mathcal{N}\left(F\left(G(u)_{j}\right)_{i},\sigma_{1}^{2}\right)$,
where the mean of $T$ is the solution of the direct problem. Given the likelihood (\ref{eq:likelihood}) and
the prior distribution (\ref{eq:prior2}), we obtain a model of the posterior distribution of $\Theta$ given $T$
using Bayes identity

\begin{equation}
\label{eq:posterior}
\pi_{\bf\Theta|\bf T}(\theta|T)\propto\pi_{\bf T|\bf\Theta}(T|\theta)\pi_{\bf\Theta}(\theta).
\end{equation}

Of note, no analytical expressions are available of the posterior distribution~(\ref{eq:posterior}), hence
we resort to sampling with Markov Chain Monte Carlo Methods. This is achieved by means of the
Metropolis-Hastings algorithm.

Let us denote by $q(\theta^{'}|\theta)$ a proposal for a probability transition kernel (in the numerical
examples in Section~\ref{sec:res&disc} we shall use the probability transition kernel known as {\it twalk} 
of Christen and Fox \cite{christen2010general}). Let us denote
$\theta_{m}$ a state in the parameter space and $\theta^{'}=(u^{'},\sigma^{'}_{2})$ a sample drawn from
the proposed probability transition kernel. Then, the standard Metropolis-Hastings algorithm
is given by Algorithm \ref{alg:mh}

\begin{algorithm}[htbp]
 \KwData{Initial point in the parameter space $\theta_{0}$, number of samples $N\in\mathbb{N}$}
 \KwResult{Samples $\{\theta_{1},...,\theta_{N}\}$ of the posterior distribution~(\ref{eq:posterior})}
 Initialization: Set $m=0$, $\theta_{0}$\;
 \For{$m=0,1,\cdots,N-1$}{
  draw $\theta'$ from the probability transition kernel $q(\cdot|\theta_{m})$\;
  compute
  \begin{equation}\label{SMHquotient}
  \alpha = \min\left\{1,\frac{\pi_{\bf\Theta|\bf T}(\theta'|T)q(\theta'|\theta_{m})}{\pi_{\bf\Theta|\bf T}(\theta_{m}|T)q(\theta_{m}|\theta')}\right\};
  \end{equation}

  draw $t$ from $U(0,1)$\;
  \eIf{$ \alpha >t$}{
   set $\theta_{m+1}=\theta'$\;
   }{
   set $\theta_{m+1}=\theta_{m}$\;
  }
 }
 \caption{Standard Metropolis Hastings algorithm\label{alg:mh}.}
\end{algorithm}

Of note, quotient \eqref{SMHquotient} has to be calculated through equation \eqref{eq:prior2},
which involves the unknown normalizing constant $Z_{0}(\sigma_{2})$. In order to circumvent
this problem, we modify the Single Variable Exchange algorithm described
in Nicholls {\it et al.}~\cite{nicholls2012coupled}, by means of the unbiased estimator
\begin{displaymath}
E=\frac{\pi_{\bf\Theta|\bf T}(\theta'|T)\pi_{\bf U|\bf\Sigma_{2}}(x|\sigma_{2})q(\theta'|\theta)}
{\pi_{\bf\Theta|\bf T}(\theta|T)\pi_{\bf U|\bf\Sigma_{2}}(x|\sigma_{2}')q(\theta|\theta')}
\end{displaymath}
of the ratio \eqref{SMHquotient}. Here $\theta=(u,\sigma_{2})$ and $x$ denotes a sample drawn
from the conditional probability distribution $\pi_{{\bf U|\Sigma_{2}}}(\cdot|\sigma_{2}')$. Then,
equations \eqref{eq:prior2} and~\eqref{eq:posterior} allow us to write\\

\begin{equation}
\label{eq:proposal}
\begin{split}
E&=\frac{\pi_{\bf T|\bf\Theta}(T|\theta')\frac{h(u',\sigma_{2}')}{Z_{0}(\sigma_{2}')}\pi_{\bf\Sigma_{2}}(\sigma_{2}')
\chi_{Q}(u')\frac{h(x,\sigma_{2})}{Z_{0}(\sigma_{2})}q(\theta'|\theta)}
{\pi_{\bf T|\bf\Theta}(T|\theta)\frac{h(u,\sigma_{2})}{Z_{0}(\sigma_{2})}\pi_{\bf\Sigma_{2}}(\sigma_{2})
\chi_{Q}(u)\frac{h(x,\sigma_{2}')}{Z_{0}(\sigma_{2}')}q(\theta|\theta')}\\
&=\frac{\pi_{\bf T|\bf\Theta}(T|\theta')h(u',\sigma_{2}')\pi_{\bf\Sigma_{2}}(\sigma_{2}')
\chi_{Q}(u')h(x,\sigma_{2})q(\theta'|\theta)}
{\pi_{\bf T|\bf\Theta}(T|\theta)h(u,\sigma_{2})\pi_{\bf\Sigma_{2}}(\sigma_{2})
\chi_{Q}(u)h(x,\sigma_{2}')q(\theta|\theta')}.
\end{split}
\end{equation}

Note that the unknown normalizing constants $Z_{0}(\sigma_{2})$ and $Z_{0}(\sigma_{2}')$ do not appear
in equation~(\ref{eq:proposal}) and the quotient is well defined if $u\in Q$. This gives rise to
Algorithm~\ref{alg:sea}

\begin{algorithm}[htbp]
 \KwData{Initial point $\theta_{0}=(u^{0},\sigma_{2}^{0})$ with $ u^{0}\in Q$, number of samples $N\in\mathbb{N}$}
 \KwResult{Samples $\{\theta_{1},...,\theta_{N}\}$ of the posterior distribution~(\ref{eq:posterior})}
 Initialization: Set $m=0$, $\theta_{m}$\;
 \For{$m=0,1,\cdots,N-1$}{
  set $\theta_{m+1}=\theta_{m}$\;
  draw $\theta'=(u',\sigma_{2}')$ from the probability transition kernel $q(\cdot|\theta_{m})$\;
  \If{$u'\in Q$}{
   draw $x$ from the conditional distribution $\pi_{\bf U|\bf\Sigma_{2}}(\cdot|\sigma_{2}')$\;
   compute
   \begin{displaymath}
   \alpha = \min\left\{1,\frac{\pi_{\bf T|\bf\Theta}(T|\theta')h(u',\sigma_{2}')\pi_{\bf\Sigma_{2}}
   (\sigma_{2}')h(x,\sigma_{2}^{m})q(\theta'|\theta_{m})}
   {\pi_{\bf T|\bf\Theta}(T|\theta_{m})h(u^{m},\sigma_{2}^{m})
   \pi_{\bf\Sigma_{2}}(\sigma_{2}^{m})h(x,\sigma_{2}')q(\theta_{m}|\theta')}\right\}\;
   \end{displaymath}

   draw $t$ from $U(0,1)$\;
   \If{$ \alpha >t$}{
    set $u^{m+1}=u'$ and $\sigma_{2}^{m+1}=\sigma_{2}'$\;
   }
  }
 }
 \caption{Single Variable Exchange Metropolis Hastings algorithm\label{alg:sea}.}
\end{algorithm}

Algorithm~\ref{alg:sea} belongs to a wide class of Metropolis-Hastings algorithms with randomized
acceptance probability discussed by Nicholls {\it et al.}~\cite{nicholls2012coupled}, where the ratio of
two target distributions is not available, but the ratio can be approximated by an unbiased estimator,
e.g.  $E$ in equation~(\ref{eq:proposal}). The equilibrium distribution of the modified algorithm is the
original target distribution.

\section{Numerical experiments}
\label{sec:res&disc}

In the first part of this Section we explore numerically how the forward mapping propagates uncertainty
in the thermal conductivity coefficent. Later, we explore the posterior distribution~(\ref{eq:posterior}) using
Algorithm~\ref{alg:sea} in two numerical examples. Hypotheses {\bf (H1)}-{\bf (H5)} are assumed to hold
in all cases. We have used Tylose parameters, a setup of parameter values and
dimensions described in Table~\ref{tab:pars}. This parameter set has already been used as a valid synthetic
setting in \cite{Smith2014}. We have solved the direct problem~\eqref{eq:pb_unid_1} approximating the polar
Laplacian and the derivatives in the boundary conditions by means of standard centered order two finite
differences and using the classical Crank-Nicolson time stepping, resulting in a second order method, both in
time and space. We have discretized the spatial domain
with 102 grid points, and we have taken 350 timesteps. Synthetic data is created solving the same direct problem
with a grid 100 times more fine in time. Observations of temperature at both, the center and a point of the boundary of the ball, are assumed to be poluted with noise according to likelihood \eqref{eq:likelihood}, where we have chosen the noise standard deviation $\sigma_{1}$ such that the signal to noise ratio (SNR) satisfies

\begin{equation}
\label{eq:snr}
\text{SNR}=\frac{\text{mean}(T)}{\sigma_{1}}=10^{3},
\end{equation}
and $\text{mean}(T)$ is the mean of the temperature $T=T(r,t)$ for $(t,r)\in B_{R}\times (0,t_{\rm f})$
measured in Kelvin degrees.
We remark that thermocouples used in real applications to aquire temperature data have roughly the
same signal to noise ratio~\eqref{eq:snr}.

\begin{table}[ht!]
\centering
\resizebox{1.\textwidth}{!}{\begin{minipage}{\textwidth}
\begin{tabular}{lp{3cm}cp{2cm}cp{2cm}}
Name  & symbol & value and dimension \\
 \hline
Thermal expansion coefficient  & $\alpha$ & $4.217\times 10^{-4}\,{\rm K}^{-1}$ \\
Density & $\varrho$ & $1000.6\,{\rm kg}\,{\rm m}^{-3}$\\
Specific heat & $C_{p}$ & $3780\,{\rm J}\,{\rm kg}^{-1}{\rm K}^{-1}$\\
Heat exchange coefficient & $h$ & $28\,{\rm W}\,{\rm m}^{-2}{\rm K}^{-1}$\\
Inner radius  & $r_{0}$ & $0\,{\rm m}$\\
Outer radius & $R$ &  $0.045\,{\rm m}$\\
Initial temperature & $T_0$ & $295\,{\rm K}$\\
Pressure increase rate & $\beta$ & $120/61\times 10^{6}\,{\rm Pa}\,{\rm s}^{-1}$\\
Final time  & $t_{\rm f}$ & $1000\,{\rm s}$\\
 \hline
\end{tabular}
\caption{{\bf Parameter setup}. We have used parameters and dimensions as described in
~\cite{Smith2014}}
\label{tab:pars}
\end{minipage}}
\end{table}

\subsection{Sensitivity analysis}

\begin{figure}
\centering
\resizebox*{13cm}{!}{\includegraphics{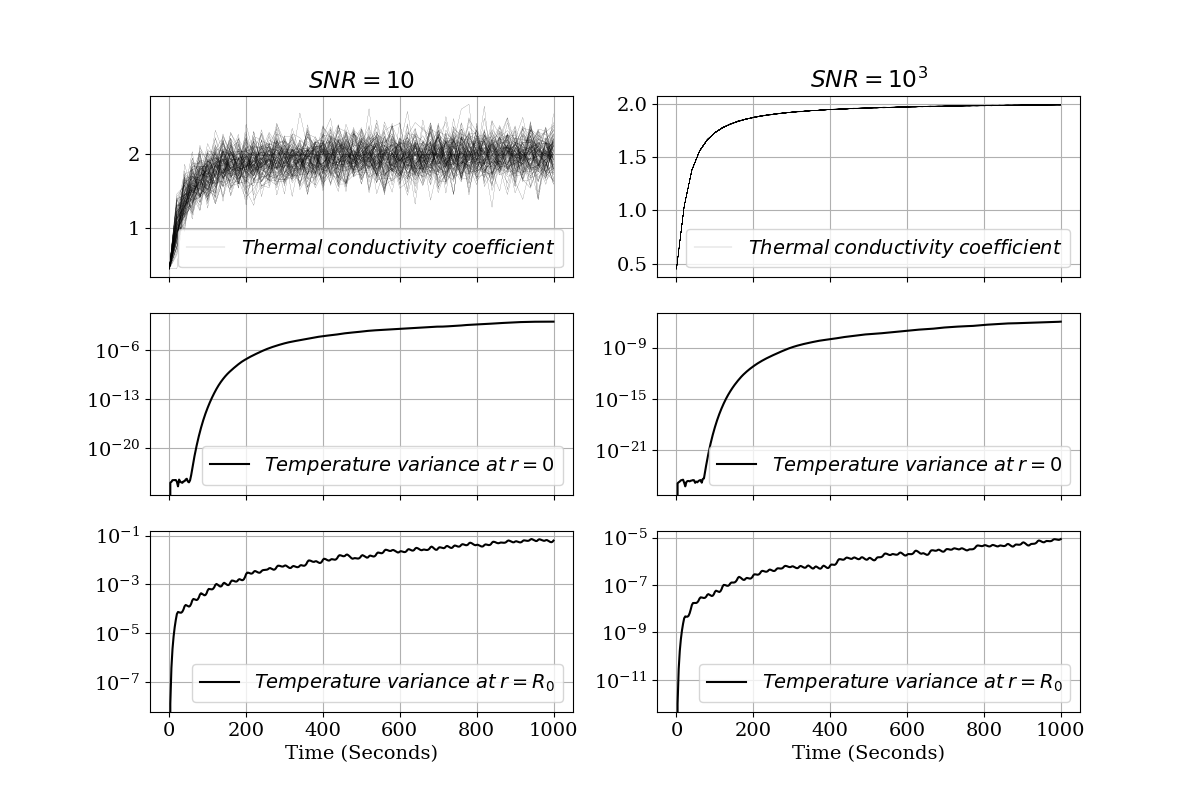}}
\caption{{\bf Uncertainty propagation.} Subplots in the top row depict 100 perturbed samples of the 
thermal conductivity coeffcient, given by equation~\eqref{eq:example1}, for $SNR=10$ and 
$SNR=10^{3}$ respectively. Subplots in the middle and the bottom rows depict the variance 
of the numerical simulation of forward mapping~\eqref{eq:forward_mapping} acting on the above 
conductivity coeffcients at $r=0$ and $r=R$ respectively. The smoothing nature of the forward mapping 
makes it necessary to aquire temperature data at large integration times, e.g. 1000 seconds.}
\label{fig:unc_prop}
\end{figure}

Figure~\ref{fig:unc_prop} illustrates that perturbations in the thermal conductivity coefficient
$k=k(t)$, given by equation~\eqref{eq:example1}, lead to rather small variance in the corresponding
temperature $T=T(r,t)$ at $r=0$ and $r=R$. Left column has signal to noise ratio $SNR=10$ and
right column has signal to noise ration $SNR=10^3$. This feature of
the forward mapping~\eqref{eq:forward_mapping} pinpoints how difficult it is to solve the corresponding
inverse problem. Indeed, as Kaipio and Fox~\cite{kaipio2011bayesian} establish, such a narrow posterior
distribution
might give rise to a situation where a computational forward mapping leads to an unfeasible posterior
model, e.g., a posterior model such that the true thermal conductivity coefficient has arbitrarily small
probability. Of note,
Figure~\ref{fig:unc_prop} also illustrates that the temperature variance is an increasing function of time.
Therefore it is possible to design observation times large enough to circumvent the smoothing effect
of the forward problem. In the examples shown below we have chosen final time $t_f=1000$ s.

\subsection{Parameter estimation}
In each case we have carried out $5\times10^{5}$ steps of the Markov Chain Monte Carlo. In Algorithm
\ref{alg:sea}, we have used the proposal $q(\theta^{'}|\theta)$ given by the {\it twalk} of Christen
and Fox \cite{christen2010general}. Our choice was based on the facts that the {\it twalk} proposal
has no tunning parameters and commutes with affine transformations of space. All programs for the
numerical experiments below were coded in python 3.7 and are available
in {\it github} (https://github.com/MarcosACapistran/diffusion)
Following standard notation, we denote by $\theta_{MAP}$ the maximum a posteriori probability estimate 
and by $\theta_{CM}$ the conditional mean estimate.

\begin{description}
\item[Example 1] In the first synthetic example we have used parameters described in
Table \ref{tab:pars} and thermal conductivity is given by the function
\begin{equation}
\label{eq:example1}
k(t)=\arctan{\frac{t}{30}}+ 0.45.
\end{equation}

\item[Example 2] In the second synthetic example, we have kept the same parameters from
the previous example, except the thermal conductivity, taking in this case
\begin{equation}
\label{eq:example2}
k(t)=\frac{1}{2}\left(\arctan{\frac{t-450}{150}}+3.5\right).
\end{equation}
\end{description}

\begin{figure}
\centering
\subfloat[\label{fig:trace1} Trace plot]{\resizebox*{5cm}{!}{\includegraphics[]{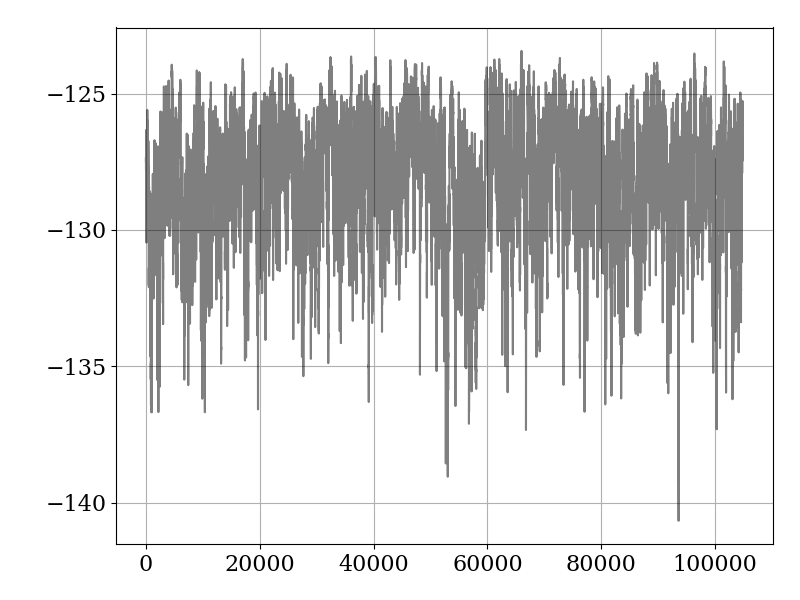}}}\hspace{5pt} 
\subfloat[\label{fig:hyperpar1} Posterior distribution of $\sigma_2$]{\resizebox*{5cm}{!}{\includegraphics[]{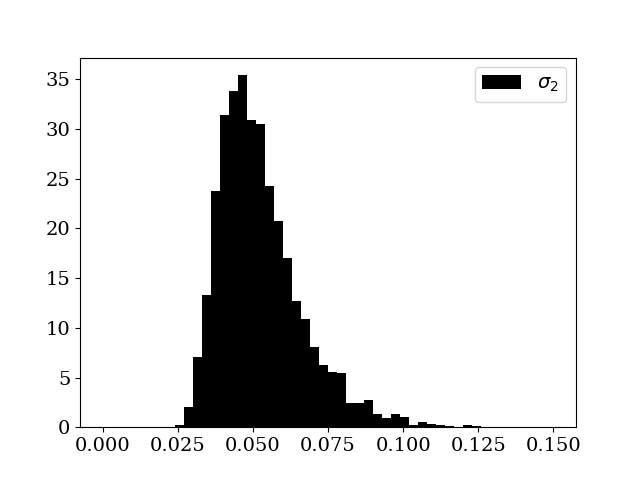}}}\\   
\subfloat[\label{fig:estimators1}{\bf True and estimators} Middle row depicts 100 samples of the posterior
distribution of the thermal conductivity coefficient. At the bottom row are shown the corresponding temperatures 
evaluated at $r=0$ and $r=R$ respectively. Hierarchical modeling ]{\resizebox*{10cm}{!}{\includegraphics[]{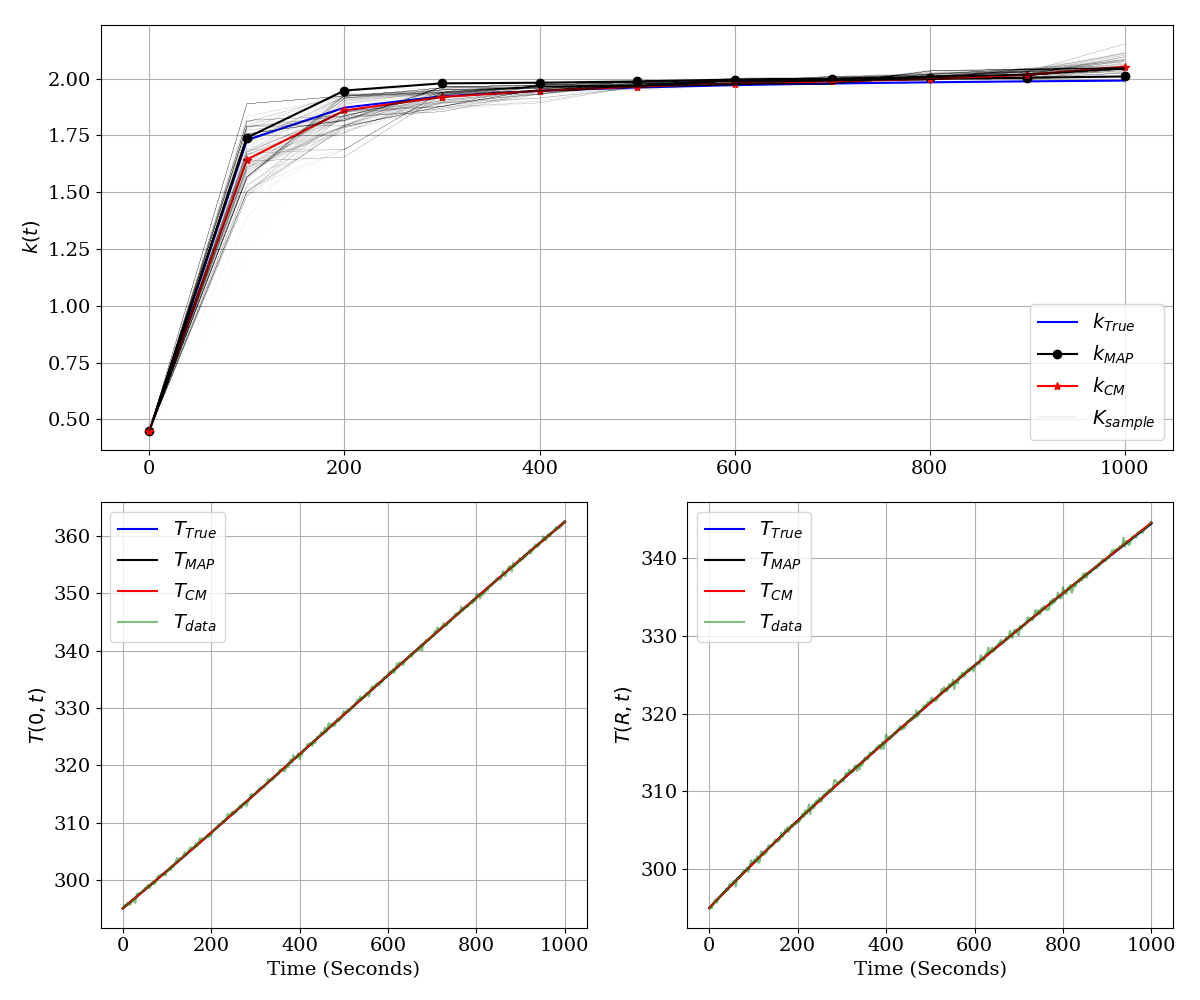}}}   
\caption{{\bf Example 1.} Subplot (a) shows }
\label{fig:example1}
\end{figure}

\begin{figure}
\centering
\subfloat[\label{fig:trace2} Trace plot]{\resizebox*{5cm}{!}{\includegraphics[]{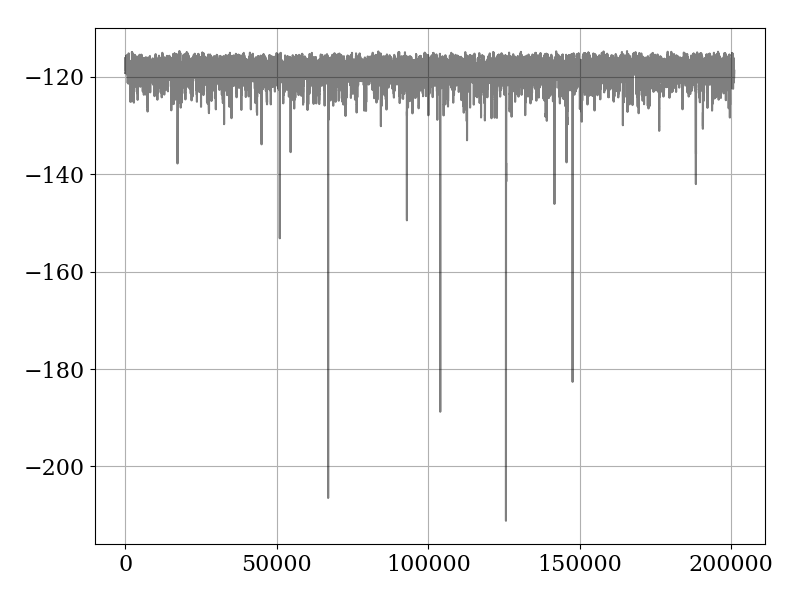}}}\hspace{5pt} 
\subfloat[\label{fig:hyperpar2}Posterior distribution of $\sigma_2$]{\resizebox*{5cm}{!}{\includegraphics[]{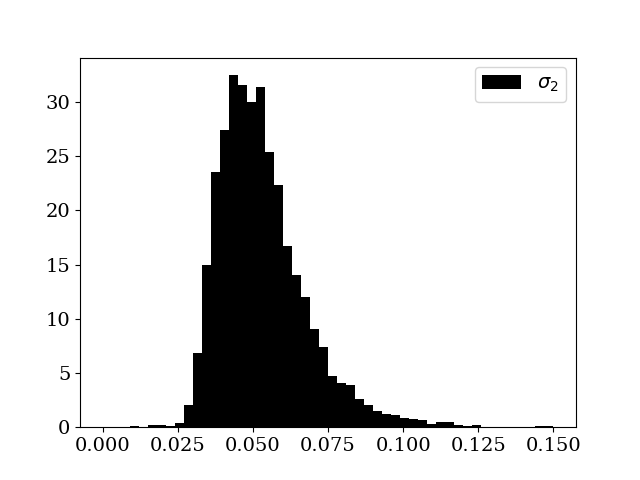}}}\\   
\subfloat[\label{fig:estimators2}Truth and estimators]{\resizebox*{10cm}{!}{\includegraphics[]{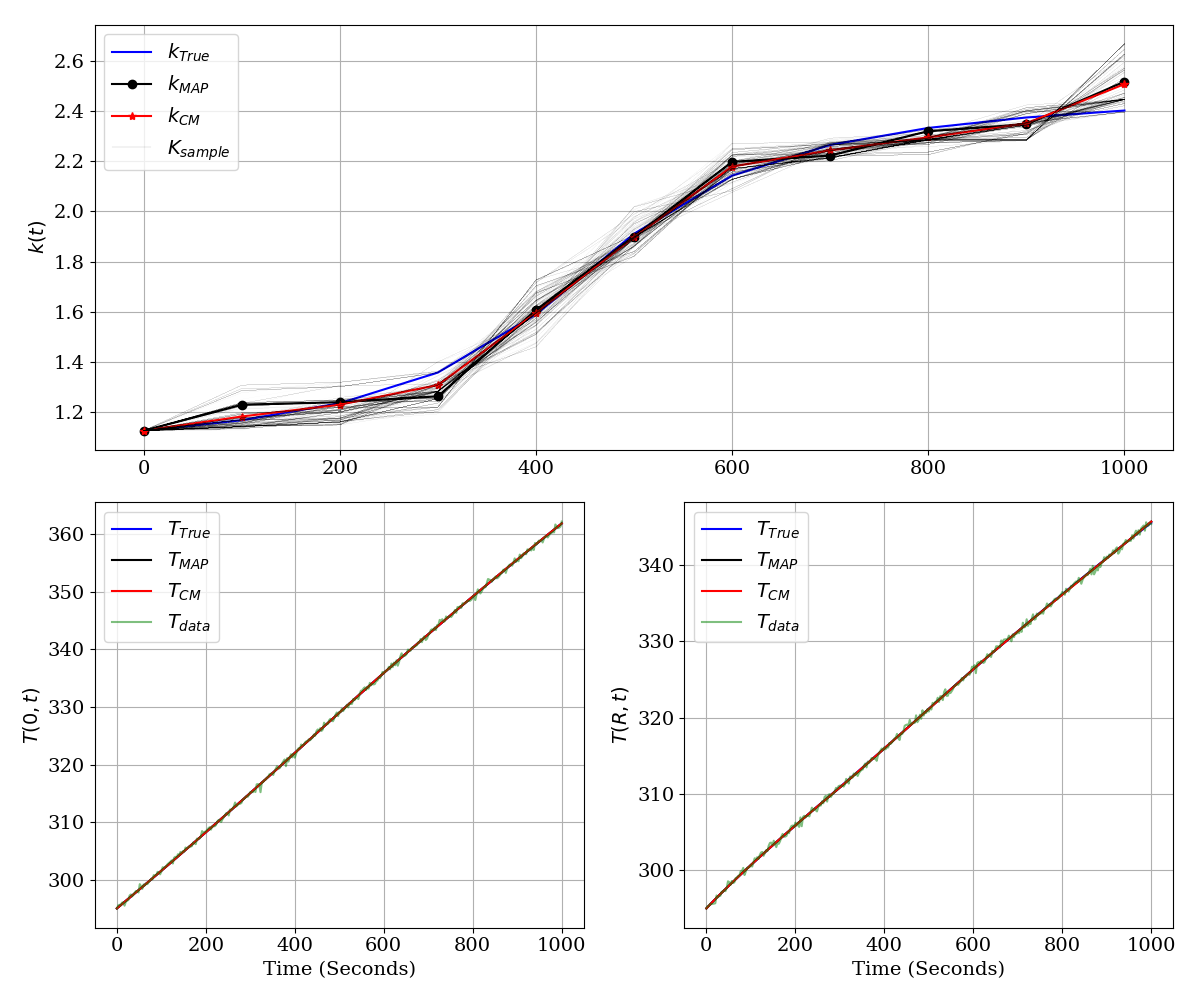}}}   
\caption{{\bf Example 2.} Subplot (a) shows }
\label{fig:example2}
\end{figure}

Figures \ref{fig:example1} and \ref{fig:example2} show the results of
identyfing the thermal conductivity coefficient $k(t)$ for examples 1 and 2 respectively.
Figures~\ref{fig:trace1} and~\ref{fig:trace2} are numerical evidence of the convergence of the Markov
Chain Monte Carlo in both examples. We have taken $SNR =10^{3}$. Of note, Figures~\ref{fig:hyperpar1} and~\ref{fig:hyperpar2} show roughly the same marginal posterior distribution for the variance
of the prior model for both examples. Finally, Figures~\ref{fig:estimators1} and~\ref{fig:estimators2}
provide further evidence that of the smoothing property of the numerical simulation of the forward
mapping~\eqref{eq:forward_mapping}. Although we have used a low dimensional 
representation~\eqref{eq:kexpu}-\eqref{eq:discrete_k} with $n=10$ in both cases, 
it is apparent that the true value of the of the thermal conductivity coeffcient lies in the support of the
posterior model.

\section{Conclusions}
\label{sec:concl}

In this paper we have shown a method to overcome the typical narrowness of the posterior
distribution arising in an inverse heat diffusion problem with applications in food technology. 
Our strategy is to construct a hierarchical prior model of the thermal conductivity coefficient 
restricted to uniqueness conditions for the solution of the inverse problem. An important feature
of our apprach is that the variance of the prior model is a parameter to be inferred from data.
Finally, we propose a Single Variable Exchange Metropolis-Hastings algorithm to correctly sample
the arising posterior distribution. Numerical evidence indicates that the resulting posterior model of 
the thermal conductivity coeffcient contains the true value.

In the numerical implementation, we have resorted to approximation methods to turn the inference
of the thermal conductivity coefficient into a parametric problem. We have used a piecewise
linear function to approximate the quantity of interest, and inference is carried out over a
finite number of real coefficients.

In the narrative of food technology, the results obtained in this paper might serve as a basis to
explore the dimension of the data informed subspace as a function of the signal to noise ratio.
The reliable solution of the inverse problem with quantifiable uncertainty provides a
basis for industrial analysis towards better controled food preservation methods.

\paragraph*{Acknowledgements}
M. A. C. acknowledges partially founding from ONRG, RDECOM and CONACYT CB-2017-284451 grants. J. A. I. R. acknowledges the financial support of the Spanish Ministry of Economy and Competitiveness under the project MTM2015-64865-P and the research group MOMAT (Ref.910480) supported by {\it Universidad Complutense de Madrid}.

%\bibliographystyle{abbrv}
%\bibliography{dif_coef}

\begin{thebibliography}{10}

\bibitem{adams1991enzyme}
J.~Adams.
\newblock Enzyme inactivation during heat processing of food-stuffs.
\newblock {\em International journal of food science \& technology},
  26(1):1--20, 1991.

\bibitem{bardsley2013gaussian}
J.~M. Bardsley.
\newblock Gaussian markov random field priors for inverse problems.
\newblock {\em Inverse Problems and Imaging}, 7(2):397--416, 2013.

\bibitem{cardoso1978new}
J.~Cardoso and A.~Emery.
\newblock A new model to describe enzyme inactivation.
\newblock {\em Biotechnology and bioengineering}, 20(9):1471--1477, 1978.

\bibitem{christen2010general}
J.~A. Christen, C.~Fox, et~al.
\newblock A general purpose sampling algorithm for continuous distributions
  (the t-walk).
\newblock {\em Bayesian Analysis}, 5(2):263--281, 2010.

\bibitem{denys2000modeling}
S.~Denys, A.~M. Van~Loey, and M.~E. Hendrickx.
\newblock A modeling approach for evaluating process uniformity during batch
  high hydrostatic pressure processing: combination of a numerical heat
  transfer model and enzyme inactivation kinetics.
\newblock {\em Innovative Food Science \& Emerging Technologies}, 1(1):5--19,
  2000.

\bibitem{fraguela2013uniqueness}
A.~Fraguela, J.~Infante, A.~Ramos, and J.~Rey.
\newblock A uniqueness result for the identification of a time-dependent
  diffusion coefficient.
\newblock {\em Inverse Problems}, 29(12):125009, 2013.

\bibitem{hendrickx1998effects}
M.~Hendrickx, L.~Ludikhuyze, I.~Van~den Broeck, and C.~Weemaes.
\newblock Effects of high pressure on enzymes related to food quality.
\newblock {\em Trends in Food Science \& Technology}, 9(5):197--203, 1998.

\bibitem{infante2015identification}
J.~Infante, M.~Molina-Rodr{\'\i}guez, and {\'A}.~Ramos.
\newblock On the identification of a thermal expansion coefficient.
\newblock {\em Inverse Problems in Science and Engineering}, 23(8):1405--1424,
  2015.

\bibitem{infante2009modelling}
J.~A. Infante, B.~Ivorra, A.~Ramos, and J.~M. Rey.
\newblock On the modelling and simulation of high pressure processes and
  inactivation of enzymes in food engineering.
\newblock {\em Mathematical Models and Methods in Applied Sciences},
  19(12):2203--2229, 2009.

\bibitem{isakov2006inverse}
V.~Isakov.
\newblock {\em Inverse problems for partial differential equations}, volume
  127.
\newblock Springer, 2006.

\bibitem{kaipio2011bayesian}
J.~P. Kaipio and C.~Fox.
\newblock The bayesian framework for inverse problems in heat transfer.
\newblock {\em Heat Transfer Engineering}, 32(9):718--753, 2011.

\bibitem{nicholls2012coupled}
G.~K. Nicholls, C.~Fox, and A.~M. Watt.
\newblock Coupled mcmc with a randomized acceptance probability.
\newblock {\em arXiv preprint arXiv:1205.6857}, 2012.

\bibitem{norton2008recent}
T.~Norton and D.-W. Sun.
\newblock Recent advances in the use of high pressure as an effective
  processing technique in the food industry.
\newblock {\em Food and Bioprocess Technology}, 1(1):2--34, 2008.

\bibitem{otero2010modeling}
L.~Otero, B.~Guignon, C.~Aparicio, and P.~Sanz.
\newblock Modeling thermophysical properties of food under high pressure.
\newblock {\em Critical reviews in food science and nutrition}, 50(4):344--368,
  2010.

\bibitem{otero2007model}
L.~Otero, A.~Ramos, C.~De~Elvira, and P.~Sanz.
\newblock A model to design high-pressure processes towards an uniform
  temperature distribution.
\newblock {\em Journal of Food Engineering}, 78(4):1463--1470, 2007.

\bibitem{serment2014high}
V.~Serment-Moreno, G.~Barbosa-C{\'a}novas, J.~A. Torres, and J.~Welti-Chanes.
\newblock High-pressure processing: kinetic models for microbial and enzyme
  inactivation.
\newblock {\em Food Engineering Reviews}, 6(3):56--88, 2014.

\bibitem{Smith2014}
N.~Smith, S.~Mitchell, and A.~Ramos.
\newblock Analysis and simplification of a mathematical model for high-pressure
  food processes.
\newblock {\em Applied Mathematics and Computation}, 226:20 -- 37, 2014.

\bibitem{wang2004bayesian}
J.~Wang and N.~Zabaras.
\newblock A bayesian inference approach to the inverse heat conduction problem.
\newblock {\em International Journal of Heat and Mass Transfer},
  47(17-18):3927--3941, 2004.

\bibitem{wang2004hierarchical}
J.~Wang and N.~Zabaras.
\newblock Hierarchical bayesian models for inverse problems in heat conduction.
\newblock {\em Inverse Problems}, 21(1):183, 2004.

\bibitem{zellner1988optimal}
A.~Zellner.
\newblock Optimal information processing and bayes's theorem.
\newblock {\em The American Statistician}, 42(4):278--280, 1988.

\end{thebibliography}

\end{document}